# Disruption of Thermally-Stable Nanoscale Grain Structures by Strain Localization

**Amirhossein Khalajhedayati [a], Timothy J. Rupert [a,b,*]**

[a] Department of Chemical Engineering and Materials Science, University of California, Irvine, CA 92697, USA

[b] Department of Mechanical and Aerospace Engineering, University of California, Irvine, CA 92697, USA

[*] Email Address: trupert@uci.edu (T.J. Rupert).

Nanocrystalline metals with average grain sizes of only a few nanometers have recently been observed to fail through the formation of shear bands. Here, we investigate this phenomenon in nanocrystalline Ni which has had its grain structure stabilized by doping with W, with a specific focus on understanding how strain localization drives evolution of the nanoscale grain structure. Shear banding was initiated with both microcompression and nanoindentation experiments, followed by site-specific transmission electron microscopy to characterize the microstructure. Grain growth and texture formation were observed inside the shear bands, which had a wide variety of thicknesses. These evolved regions have well-defined edges, which rules out local temperature rise as a possible formation mechanism. No structural evolution was found in areas away from the shear bands, even in locations where significant plastic deformation had occurred, showing that plastic strain alone is not enough to cause evolution. Rather, intense strain localization is needed to induce mechanically-driven grain growth in a thermally-stable nanocrystalline alloy.

In recent years, studies have shown that plastic deformation is very different in nanocrystalline metals when compared to traditional microcrystalline materials[1]. In conventional metallurgical theory, grain boundaries are seen as obstacles to intergranular dislocation motion. However, recent studies have shown that the grain refinement to nanometer scale shifts the sources of plastic deformation from inside the crystals to the interfacial regions themselves[2,3]. Dislocations created and subsequently absorbed at grain boundaries become the main source of plastic deformation in nanocrystalline metals with average grain sizes ($d$) between 100 nm and 10 nm. Further refinement of the grains below 10 nm confines plasticity almost entirely to the grain boundaries, with grain boundary migration and grain rotation/sliding as the major events that contribute to plastic deformation[4].

As a result of these novel deformation mechanisms, Schuh and Lund[5] suggested that nanocrystalline systems with $d < 10$ nm should have mechanical properties which resemble those of metallic glasses. This concept was first supported by the observation that nanocrystalline metals exhibit a tension-compression asymmetry of their strength and shear transformation zones operate inside of grain boundary regions during plasticity[6]. Recent molecular dynamics (MD) simulations[7] have uncovered a universal relation between strength and modulus in nanocrystalline alloys that resembles prior observations of metallic glasses[8]. Another connection between the mechanical behavior of metallic glasses and nanocrystalline metals was found when Trelewicz and Schuh[9] explored the plasticity of nanocrystalline Ni-W using nanoindentation and found shear steps in the pileup surrounding the indents for average grain sizes of 6 nm and 3 nm, with more significant steps observed in the finer grained sample. The shear offsets observed in this study were similar to observations from the metallic glass literature, such as the shear banding found after nanoindentation of an Al-based metallic glass by Jiang et al.[10]. In a more recent study of nanocrystalline Ni-W, Khalajhedayati and Rupert[11] observed that shear banding is more prominent during uniaxial compression, which does not impose the confining hydrostatic pressure found in nanoindentation. In this case, shear banding was found in grain sizes as large as 15 nm, suggesting that catastrophic strain localization can be a more common problem for nanocrystalline materials than previously thought and that this phenomenon must be more thoroughly studied.

When shear banding is observed in metallic glasses, it is often accompanied by structural evolution, most commonly in the form of crystallite formation[12,13]. However, the effect of shear banding on the structure of nanocrystalline materials has not been explored thoroughly, with experimental evidence especially scarce. Wei and coworkers did observe grain coarsening inside of shear bands in ultra-fine grained (typically meaning 100 nm < $d$ < 1000 nm) Fe with $d$ of 110 nm [14], as well as ultra-fine grained W[15] and Ta[16]. However, these materials have average grain sizes that are larger than what would be considered nanocrystalline and were all body centered cubic (BCC) metals, meaning a direct extrapolation to nanocrystalline face centered cubic (FCC) metals such as Cu, Al, or Ni is not possible. Some insight has been provided by MD simulations, but these studies have been limited to pure, single element metals. For example, Sansoz and Dupant[17] studied nanoindentation of nanocrystalline Al with $d = 5$ nm and $d = 7$ nm at absolute zero temperature and reported grain growth due to grain rotation and grain boundary migration inside of shear bands. More recently, Rupert[18] observed the formation of shear bands during simulated tensile testing of a nanocrystalline Ni with $d = 3$ and 6 nm using MD. These studies show that the grain structure in pure nanocrystalline materials can be considered dynamic under mechanical loading.

However, alloying with a second element is often required in practice, since pure nanocrystalline metals can undergo thermally-driven grain growth even at room temperature in some cases[19]. Chookajorn et al.[20] showed that reducing grain boundary energy by doping these interfaces can be a powerful design strategy for the creation of thermally-stable nanocrystalline materials, demonstrating this concept by producing a W-20 at.% Ti alloy that retained a 20 nm grain size even after being exposed to 1000 °C for one week. An exemplary material system that uses this concept of grain boundary dopants to its advantage is nanocrystalline Ni-W. Detor and Schuh[21] created thermally-stable Ni-W alloys with grain sizes ranging from ~140 nm down to ~3 nm, tuned by controlling the W concentration, using electrodeposition. Recent studies have shown that alloying can also change a nanocrystalline materials resistance to mechanically-driven structural evolution. For example, extremely pure nanocrystalline Al was observed to experience dramatic grain coarsening under uniaxial loading[22], but the addition of small amounts of O into such films stabilized the grain structure[23]. Similarly, Lohmiller et al.[24] found that pure Au is prone to stress-induced grain growth, while Au with Cu segregated to the grain boundaries is not. From these studies, it is clear that doping can increase the stability of nanocrystalline grain structures, at least to evolution caused by temperature or simple stress states. However, the impact of the extreme deformation conditions which occur during shear banding has not been explored. Specifically, it would be useful to understand if the extreme deformation that occurs during shear banding can overcome the enhanced structural stability brought by alloying additions and if teven heavily doped grain structures can evolve under mechanical loading.

In this study, we investigate the microstructure of nanocrystalline Ni-W with $d$ = 5 nm before and after catastrophic shear banding. Ni-W was chosen as a model system for this study because, due to the subtle tendency for segregation of W to the grain boundaries (grain boundary excesses of 4-8 at.% W over the grain interior composition), nanocrystalline coatings that are resistant to thermal grain growth up to ~500 °C (i.e., ~45% of the melting temperature) can be created[25]. This allows us to observe how severe strain localization occurs in an alloy system which is typically stable. A combination of microcompression and nanoindentation were used to induce shear banding, since these two techniques allow us to explore different stress states and plastic strain distributions. During microcompression, nearly all plastic strain is limited to the shear bands, while in nanoindentation there is a combination of shear banding and homogeneous plasticity. We find that intense strain localization inside of shear bands can drive obvious coarsening and texturing, while homogeneous plasticity (or low plastic strain) leaves the grain structure unchanged. These results suggest that even thermally-stable nanocrystalline grain structures can coarsen under mechanical loading, but extreme conditions are required.

**Results**

**Microcompression of nanocrystalline Ni-W with $d$ = 5 nm.** Figure 1a shows a scanning electron microscope (SEM) image of a micropillar of nanocrystalline Ni-W after microcompression testing. The pillar experienced a large strain burst and failed catastrophically. The micropillar has a number of major surface steps corresponding to primary shear bands (red arrow), with a high density of thinner, secondary shear bands that crisscross each other also observed (white arrow). Figure 1b shows the corresponding true stress-strain curve for this pillar. Several strain bursts are observed in this curve, with two examples denoted by black arrows. The pillar compresses elastically for the majority of this test, with deviations from this behavior only coming in the form of strain bursts. Such a response is a sign that plasticity is dominated by shear banding for this testing methodology, not homogeneous plastic deformation, as a smooth microplastic region of the stress-strain curve would be found in that case.

Figure 2a presents a bright field TEM image of shear bands, one primary and one secondary, inside of the micropillar. Shear bands with thicknesses from 10 to 270 nm were observed, and most of the shear bands are in the lower end of this range. The thickest bands corresponded to locations where large, primary shear bands were found on the surface, with an example found in the center of Figure 2a and denoted by dashed white lines. A smaller, secondary shear band is found in the lower right of the same figure outlined by solid white lines. This figure also shows that grain growth is limited to the shear bands. Figure 2b shows a magnified view of the microstructure inside of the major shear band, with the largest grains in this region being ~17 nm in diameter. Figure 2c shows the cumulative distribution function of grain size inside and outside the shear bands, as compared to the as-deposited microstructure. The average grain size inside the shear bands has been increased by more than 50%, from 5 nm to about 8 nm, and the distribution of grain sizes has widened. The grain size outside of the shear bands, in the region which only experienced elastic deformation, has not changed and the grain size distribution follows the same pattern as the initial microstructure. An example of the network of shear bands found inside the sample is shown in Figure 2d. Part of a primary shear band can be seen in the top left of the image, with many secondary shear bands coming out in directions downward and to the right. The crisscross pattern seen on the surface of the pillar in the regions with a high density of secondary shear bands has a similar appearance. All shear bands observed in microcompression samples appeared as long, straight lines. To the best of our knowledge, this is the first time that the microstructure inside of a shear band has been observed experimentally in a nanocrystalline FCC metal.

Figure 3a presents a selected area electron diffraction pattern taken from one of the major shear bands. The segmented ring pattern is indicative of a textured microstructure and the specific pattern observed here is a pure shear texture often observed in FCC metals, consisting of {100} <011>, {111} <112>, and {111} <110> components[26]. Similar patterns and textures were found along the entire length of the shear band, confirming that this observation was a general one. A dark field image, taken by putting the smallest objective aperture on one of the segmented rings, is shown in Figure 3b. Grains that are oriented in the same diffraction condition are visible as bright spots in this image, with the high density of bright regions inside the shear band indicating the existence of a preferred orientation. Outside the shear band, the distribution of dark and bright spots is random since there is no texture. Figure 3c shows a higher magnification dark field TEM image of the microstructure inside the same shear band, where a cluster of grains with the same orientation can be seen.

**Nanoindentation of nanocrystalline Ni-W with $d = 5$ nm.** Nanoindentation imposes a strong strain gradient and a complex stress field under the indenter tip[27]. Figure 4a shows a nanoindentation made with a cube corner tip. The tip geometry literally resembles the corner of a cube and has a very sharp end, therefore it confines the plasticity to a small volume of material under the tip and often induces pile-up formation around the indents. The large shear steps that are observed in the pile-up surrounding the nanoindentation site in this figure indicate that inhomogeneous plasticity and flow localization occurred[28]. Figure 4b shows the corresponding load-displacement curve from this nanoindentation experiment. The nanoindenter is inherently force-controlled, so any shear banding events appear as a burst in displacement[11]. Two such displacement bursts are magnified in Figure 4b, each with a magnitude of ~10-15 nm.

An indentation study by Wright et al. on Zr based metallic glass showed that flow serration during loading is due to formation of shear bands under the indenter tip[29]. Although these serrations clearly exist in our experiment, careful examination shows that the loading curve is mostly smooth and only a few displacement bursts are observed. Another way to visualize the number of the shear band events is to plot the instantaneous strain rate versus indentation depth, as shown in Figure 4c. Since the indentation test was supposed to be at a constant strain rate under perfect conditions (i.e., without any displacement bursts), the strain rate data should follow a straight line. However, the displacement bursts that occur in this experiment lead to local peaks in the instantaneous strain rate. The peaks that are pointed out here correspond to the same displacement bursts that were highlighted in the load-displacement curve, and less than 10 of these peaks were observed. The deformation is a combination of homogenous plasticity (smooth load-displacement curve) and strain localization (displacement bursts). Schuh and Nieh[30] observed similar combinations of deformation modes in Pd- and Zr-based metallic glasses.

Figure 5a shows cross-sectional TEM images of the nanoindentation observed in the previous figure. The area under the indenter tip denoted by a dashed box is magnified in Figure 5b. The shear bands have a curved shape and extend deep under the indenter tip, in contrast to the straight shear bands found after microcompression. Measurement of the shear bands after indentation showed that the thickness varies from 8 nm to 32 nm. The area denoted with a solid white box is magnified in Figure 5c. Grain growth was observed in all the shear bands but, due to limited thickness of the shear bands in nanoindentation, selected area diffraction patterns could not give conclusive results about the texture. Figure 5c shows the shear bands that extend to the pile-up surrounding the indentation site. These bands terminate at the surface and result in slip steps there, as previously shown from a different perspective in Figure 4a. Figure 6a shows an example of the complex, entangled pattern of curved shear bands formed under the indent. Figure 6b presents the cumulative distribution function of grain size for the sample which was indented. The grain size distribution outside of the shear bands remained the same as the as-deposited specimen. However, the grain size inside the shear bands is much larger, with $d$ increasing from 5 nm to 8 nm and a distribution which ranged from 3-17 nm.

**Discussion**

The experiments described above show that nanocrystalline alloys with extremely fine grain sizes can experience strain localization in the form of shear banding, mimicking prior observations in metallic glasses. This similarity can be understood by recognizing the features these materials have in common. Both metallic glasses and nanocrystalline metals are extremely strong, which has been shown to be important for strain localization[3,31]. If the ratio of strain hardening rate to the applied stress drops below a critical value, a material can be prone to flow localization[32]. In both nanocrystalline metals and metallic glasses, the applied stresses are extremely high and strain hardening is almost nonexistent, satisfying these conditions. In addition, both classes of materials are unable to store large numbers of dislocations and achieve large uniform elongations. For a metallic glass, dislocations do not exist at all due to the lack of long range crystalline order. For nanocrystalline metals, dislocations can carry some fraction of the overall plastic strain but they are absorbed after traversing grains and leave no residual dislocation density. In addition, the smallest nanocrystalline grains deform through direct rotation or sliding. While the existence of shear banding in both materials and the similarities between their overall mechanical behavior is important, the evolution of grain structure is our main finding in this paper.

Our observations of structural evolution inside of nanocrystalline shear bands show that intense strain localization leads to (1) grain growth and (2) the formation of a shear texture. The first type of evolution can be understood as the material being driven closer to an equilibrium state by high stresses during deformation. Atoms in the grain boundaries are defects and exist in a higher energy state than atoms found in the grain interiors. Grain growth reduces the volume fraction of material located inside of grain boundaries, although the extent of grain growth will be a complex convolution of effects relating to the active deformation mechanisms and the driving force for growth. For a starting grain size of 5 nm, plasticity should be dominated by grain rotation and sliding, as well as grain boundary migration[33,34,35]. These mechanisms are able to efficiently coarsen grain structures, with grain rotation capable of causing grain coalescence[36,37] and boundary migration directly increasing the size of one grain at the expense of another. Rupert[18] used MD simulations to demonstrate that such mechanisms can cause grain growth during strain localization in polycrystalline nanowires with grain sizes similar to those studied here. However, as grains grow, there will be a gradual shift toward dislocation-based plasticity. Vo et al.[38] found that dislocation-based plasticity become the majority carrier of plastic strain as grains become larger than ~10 nm. Therefore, the grain growth will effectively shut itself off once grains are too large for collective grain boundary mechanisms to dominate plasticity. This idea is supported by our observation that the maximum grain size stayed below 20 nm in both nanoindentation and microcompression experiments. Another contributing factor could be supersaturation of the grain boundaries. Previous work by Detor and Schuh[21] demonstrated that only ~17-18 at.% W is enough to stabilize a $d = 8$ nm microstructure in nanocrystalline Ni-W, while our alloy here has 23 at.% W for this same grain size after coarsening. Considering that the increase in grain size has decreased the volume fraction of grain boundaries, and our careful examination of SAED patterns of shear bands shows no evidence of any second phase segregation or formation of W particles[39], the concentration of W at the grain boundaries should be even higher than the minimum value needed for thermal stability. With additional grain boundary doping, there will be a further decrease in the grain boundary energy, reducing the driving force for coarsening of the nanocrystalline structure.

Microstructural evolution toward a lower energy state is also observed in metallic glasses, where nanocrystalline structures are sometimes found inside or in the vicinity of shear bands[40]. For example, Jiang et al.[10,41] and Kim et al.[42] saw that small crystallites nucleated in indented regions of Al- and Zr-based bulk metallic glasses, respectively. Zheng et al.[43] also observed shear bands with nanocrystals inside of a MgCuGd metallic glass consolidated with high pressure torsion (HPT), where a material is deformed in a torsional mode while under a confining pressure. During shear banding, the atoms which were originally in an amorphous configuration have the ability to settle into a crystalline state. Most studies suggest that this evolution process is related to the creation of excess free volume, enhanced atomic shuffling, and high local strain rates inside of the shear bands [40,41,44]. However, other authors have found that local temperature rise inside of shear bands can cause crystallization, and this temperature rise is often accompanied by the formation of vein patterns or liquid like droplets[45,46] on the surface of the sheared specimen. Evolution driven by local temperature fluctuations will also result in diffuse shear bands, with gradual structural evolution as one moves away from the band[40]. However, our experimental observations suggest that such local temperature rise is not a contributing factor to the structural evolution in nanocrystalline Ni-W. SEM images of the failure surfaces from the microcompression samples showed no vein patterns and the slip surfaces resembled a clean fracture. In addition, our TEM images showed shear bands with well-defined edges and no gradient in the microstructure was found in the vicinity of the shear bands. As such, it appears that our evolution is purely driven by mechanical stresses.

The observation of texture formation can be explained by a shift to dislocation-based mechanism through a combination of grain growth and high plastic strains. Previous studies[47,48] have shown that plasticity due to grain boundary sliding and grain rotation randomizes texture, similar to observations of superplastic behavior in nanocrystalline metals[49]. However, once grains get larger than ~10 nm[38], plasticity is dominated by dislocation glide, which has been shown to induce preferred orientation (see, e.g., the formation of simple shear texture in nanocrystalline Ni[50]). The dominant deformation mechanism is also a function of plastic strain, with higher plastic strains increasing the importance of dislocation-based mechanisms[38]. This concept is supported by literature data from processing experiments. For example, Ivanisenko et al.[47] employed HPT on a nanocrystalline Pd and found that extremely high shear strains of 124 can result in grain growth and texture formation. However, Markmann et al.[48] showed that cold rolling of a similar nanocrystalline Pd to much lower plastic strain values does not cause any texture formation. Skrotzki et al.[51] studied the effect of severe plastic deformation on the texture of nanocrystalline Pd-Au more systematically, finding that a random texture was observed at low applied strains. This was then replaced by a "brass-type" shear texture once the applied shear strain exceeded 1.

The effect of the overall stress state can be understood by comparing the results of the microcompression with those of the nanoindentation. While a microcompression experiment applies a simple uniaxial stress to the micropillar, the indentation stress state is much more complicated under the indenter tip, having a combination of hydrostatic and deviatoric stress components. In addition, the stress-strain curve for microcompression shows that the only deviation from elasticity is discrete plastic deformation through shear banding. As mentioned in the Results section, there is a combination of homogeneous plasticity and shear banding in the case of our nanoindentation experiments. We find that the nanoindentation curve is smooth with just less than 10 displacement bursts caused by shear bands, each only ~10-15 nm in length. Additional experiments with even lower strain rates and higher data acquisition rates, which should make it easier to activate and identify shear bands, also showed the same behavior. If shear banding was the only contributor to plastic deformation, subtracting the pop-in displacements from the load-displacement curve would give a loading curve that is predicted by Hertzian elasticity theory and negligible residual indentation depth after unloading. This is clearly not the case for our experiments on Ni-W, as a large plastic impression is left on the surface even through shear banding should only contribute ~100 nm of plastic depth.

Shear banding in nanoindentation experiments, in contrast to microcompression, cannot be responsible for all of the plastic deformation. Therefore, homogenous plasticity outside of the shear bands must have occurred during nanoindentation. However, by looking at the grain size distribution outside of the shear bands but within the plastic zone of the indentation, we find that homogenous plasticity does not drive structural evolution, as the average grain size and even the details of the grain size distribution are identical to the as-deposited microstructure. This observation contrasts with previous observations of pure nanocrystalline metals, which experience grain growth during the early stages of plasticity in uniaxial tension experiments[22]. Doping of grain boundaries with W makes the microstructure stable against coarsening at small plastic strains, analogous to the increase in thermal stability, but this structural stability can be overcome when intense shear localization occurs. These results suggest that the microstructure of even thermally-stable nanocrystalline FCC alloys should be considered dynamic, if extreme deformation conditions such as catastrophic shear banding can be accessed.

To summarize, the effects of shear banding on local grain structure were studied in a nanocrystalline Ni-W alloy with an as-deposited grain size of 5 nm. Microcompression experiments as well as nanoindentation tests were used to study the effects of different stress fields and deformation boundary conditions. Evidence of significant grain growth and texturing as a result of strain localization was found inside of the shear bands. A degree of mechanically-induced grain growth was observed for both testing conditions, suggesting an inherent limit on this coarsening behavior that depends on multiple factors such as the dominant deformation physics and the driving force for grain growth, both of which change with grain size. Our results showed that the nanocrystalline Ni-W can be susceptible to grain growth only when extreme values of shear strain are applied, and suggest that this is a purely stress-driven process and not caused by local heating inside the shear bands.

**Methods**

A Ni-based alloy with 23 at.% W and an average grain size of 5 nm was prepared using the pulsed electrodeposition method of Detor and Schuh[21]. Although the equilibrium Ni-W phase diagram predicts a mixture of FCC Ni(W) solid solution and $Ni_4W$ for this alloy composition, our films were comprised only of FCC Ni(W) solid solution, in line with previous observations of similar nanocrystalline Ni-W alloys [25,39]. Cylindrical micropillars with average diameters of 7 µm and lengths of 16 µm were fabricated using a lathe milling Focused Ion Beam (FIB) technique[11]. An Agilent G200 nanoindenter with a flat triangular tip was used to compress the pillars at a constant displacement rate of 5 nm/s. For our nanoindentation studies, a cube corner tip was used to make the observation of the deformed region around the indent easier, as significant pile-up can occur with this tip geometry[52]. Previous studies[30,53] have shown that shear banding during nanoindentation is strain rate dependent and that a higher strain rate shifts the plasticity toward more homogenous plasticity. For this reason, a relatively low strain rate of $0.05\ s^{-1}$ was chosen to promote shear banding. For nanoindentation, the as-deposited sample was annealed at 300°C for 1 hour to relax the grain boundary structure to a lower energy level[52] since relaxation has been shown to increase the propensity for shear localization. TEM characterization was performed with a Philips-FEI CM-20 operated at 200 kV and equipped with a Gatan digital camera. Site-specific TEM samples were prepared using a FIB in-situ lift out technique[54] in a FEI Quanta 3D field emission gun (FEG) dual beam microscope. The TEM lamellas were polished with a 5 kV $Ga^+$ ion beam to minimize ion beam damage to the sample. All SEM imaging was carried out at 5 kV in the same dual beam SEM/FIB.

**Acknowledgments**

This work was supported by the U.S. Army Research Office, through Grant W911NF-12-1-0511.

**Author contributions**

A.K. carried out the experimental work. All authors contributed to discussion and analysis of the data, as well as manuscript writing.

**Additional information**

Competing financial interests: The authors declare no competing financial interests

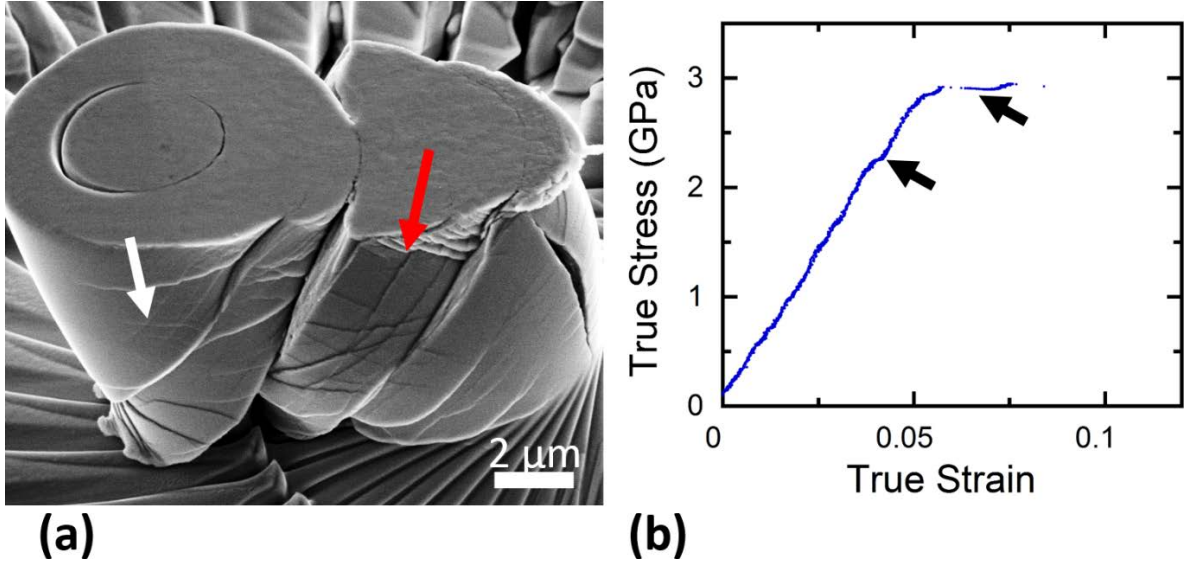

**Figure 1. Microcompression of nanocrystalline Ni-W with an average grain size of 5 nm:** (a) SEM image of a micropillar after deformation, taken at 52° tilt. Several slips steps from primary shear bands are observed on the surface, in addition to smaller, secondary shear bands criss-crossing in between. (b) True stress-strain curve of the corresponding pillar showing several displacement bursts during microcompression.

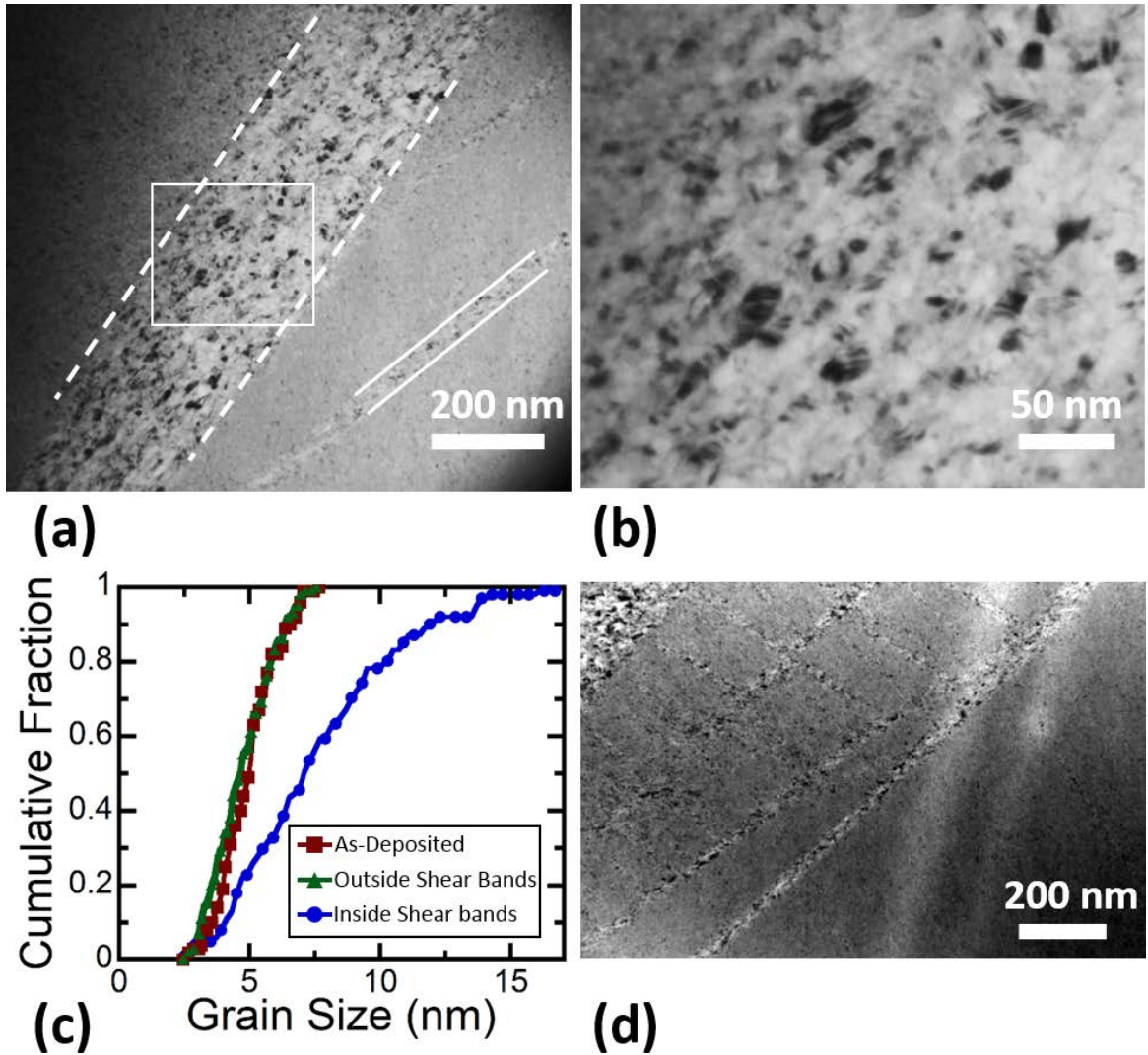

**Figure 2. Microstructure of the shear bands inside the pillar:** (a) TEM image of a primary shear band outlined with dashed white lines and secondary shear band outlined with solid white lines. (b) Magnified view of the area denoted by the white box in (a) showing grain growth inside of the shear band. (c) Cumulative distribution function of grain size from grains inside and outside the shear bands, as well as the as-deposited sample. (d) TEM image showing an example of the shear band network inside of the pillar.

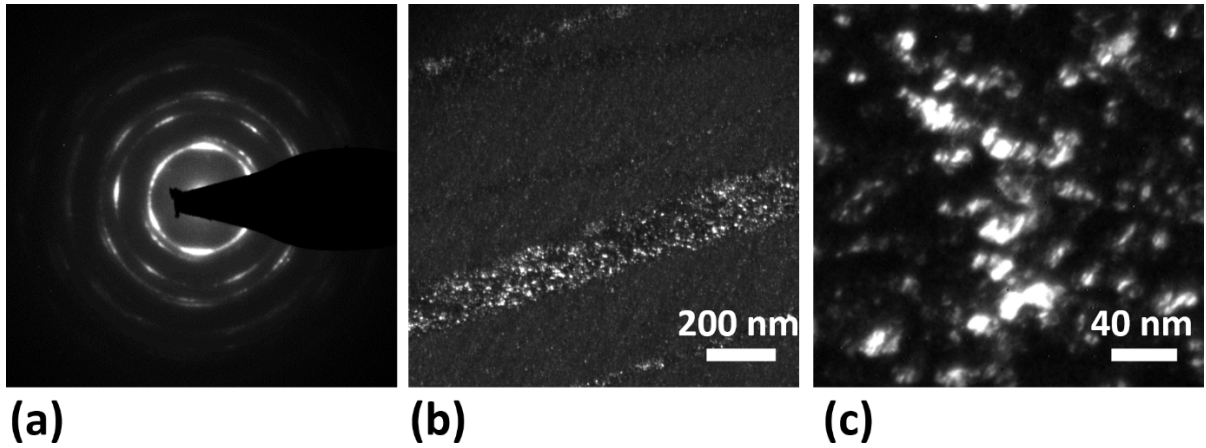

**Figure 3. Pure shear texture in the shear band:** (a) Selected area diffraction pattern taken from a thick shear band. (b) Corresponding dark field TEM image of the shear band, showing many grains with the preferred texture within the shear bands. (c) Magnified view of the microstructure inside the shear band, showing a cluster of grains with similar orientations.

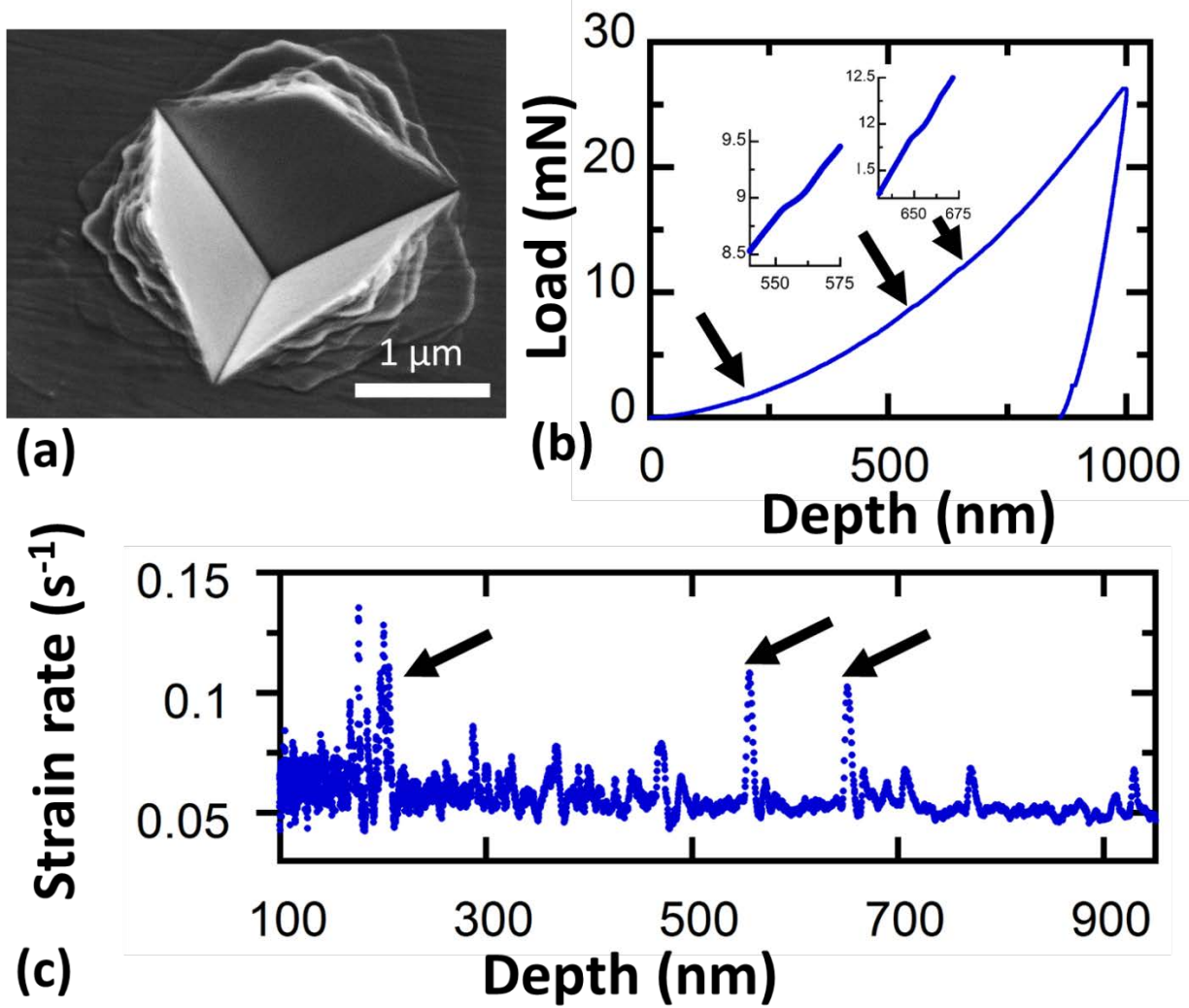

**Figure 4. Nanoindentation of nanocrystalline Ni-W with average grain size of 5 nm using a cube corner tip:** (a) SEM image of one of the indents taken at 30° tilt. (b) Corresponding nanoindentation load-displacement curve, with several strain bursts denoted by black arrows. (c) Average strain rate versus indentation depth for the same loading curve, showing jumps in the instantaneous strain rate during loading which match up with the bursts denoted in (b).

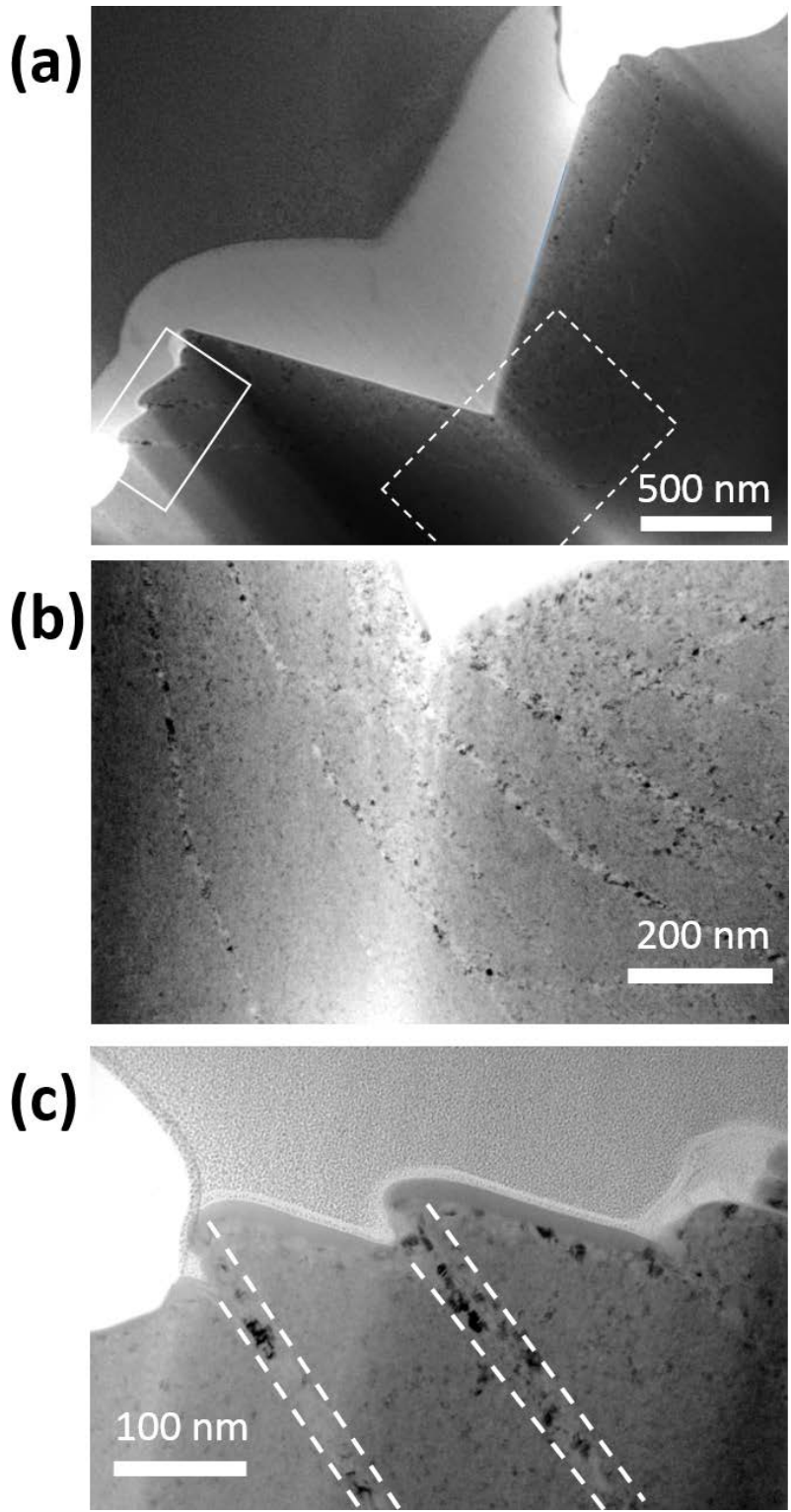

**Figure 5. TEM images of Ni-W after nanoindentation:** (a) TEM image of the entire indentation cross-section. (b) Magnified image of the dashed box showing the shear bands diretly underneath the indenter tip. (c) Magnified image of the solid box, showing shear bands which terminate at the surface and create slip steps.

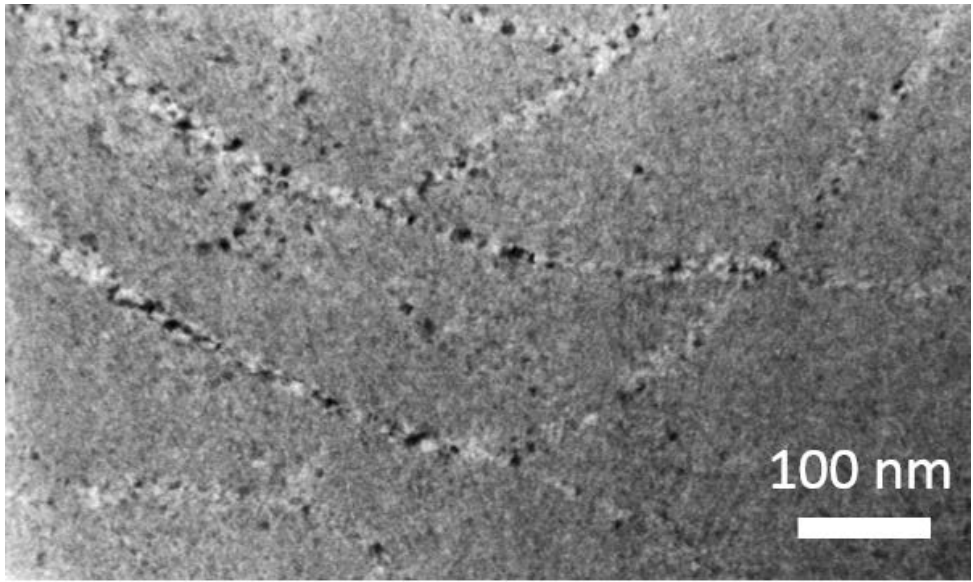

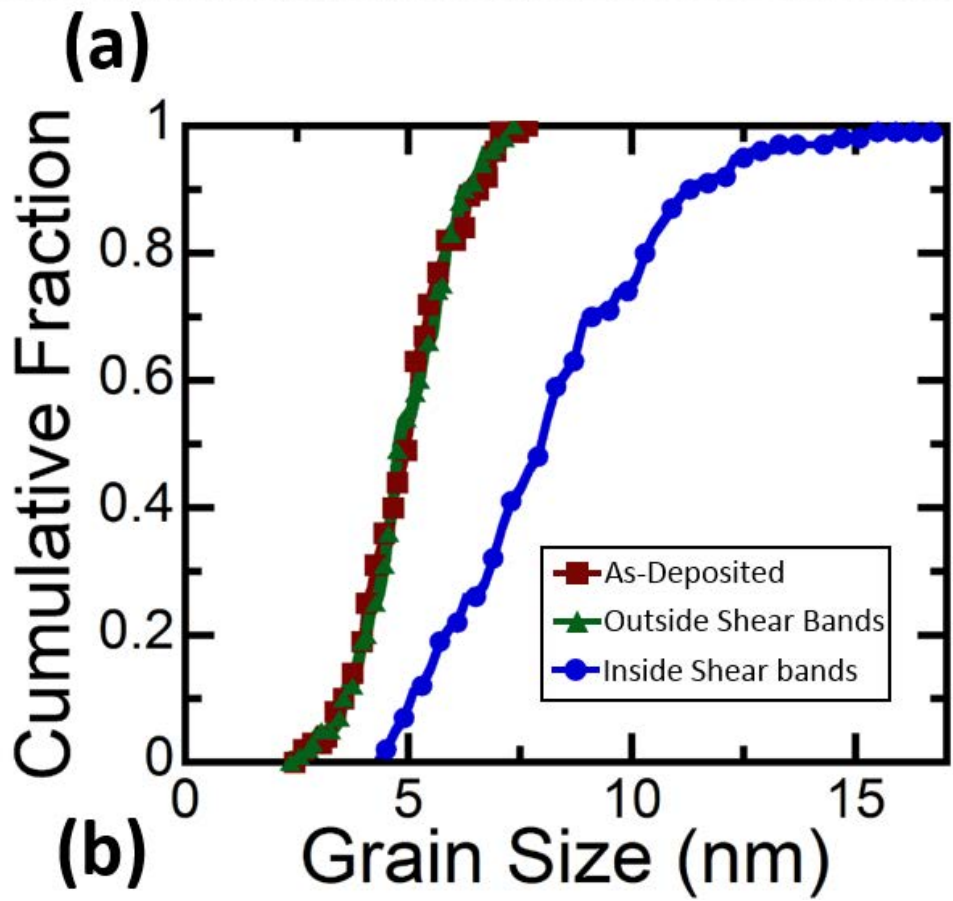

**Figure 6. Microstructure after nanoindentation:** (a) The microstructure of the same indent in the pile-up on the surface of the sample; TEM image of shear bands forming an intersecting network underneath an indent. (b) Cumulative distribution function of grain size Cumulative fraction vs. grain size from grains inside and outside the shear bands, as well as the original as-deposited relaxed microstructure.